\documentclass[12pt,preprint]{aastex631}

\usepackage{natbib}
\usepackage{graphicx}
\usepackage{url}
\usepackage{indentfirst}
\usepackage{gensymb}
\usepackage{textcomp}
\usepackage{amsmath}
\usepackage{physics}
\usepackage{listings}
\usepackage[title]{appendix}

\makeatletter
\def\@to{to}
\makeatother

\DeclareGraphicsExtensions{.png,.jpg}
\usepackage[utf8]{inputenc}

\newcommand{\Ha}{H$\alpha$}
\newcommand{\Hb}{H$\beta$}

\newcommand{\OI}{[O~{\sc i}]}

\newcommand{\NII}{[N~{\sc ii}]}

\newcommand{\SII}{[S~{\sc ii}]}

\newcommand{\OIII}{[O~{\sc iii}]}

\newcommand{\h}{$h_{70}^{-1}$}
\newcommand{\be}{\begin{equation}}
\newcommand{\ee}{\end{equation}}
\newcommand{\ba}{\begin{eqnarray}}
\newcommand{\ea}{\end{eqnarray}}

\newcommand{\simgt}{\lower 2pt \hbox{$\, \buildrel {\scriptstyle >}\over {\scriptstyle\sim}\,$}}
\newcommand{\simlt}{\lower 2pt \hbox{$\, \buildrel {\scriptstyle <}\over {\scriptstyle\sim}\,$}}

\newcommand{\ls}{\lower 2pt \hbox{$\;\scriptscriptstyle \buildrel<\over\sim\;$}}
\newcommand{\gs}{\lower 2pt \hbox{$\;\scriptscriptstyle \buildrel>\over\sim\;$}}

\def\apjl{\rmfamily{ApJ~}}        
\def\aap{\rmfamily{A\&A~}}        
\def\apjs{\rmfamily{ApJS~}}       

\begin{document}

\title{Variability Selected Active Galactic Nuclei from ASAS-SN Survey: Constraining the Low Luminosity AGN Population}
\author[0000-0002-7720-3418]{Heechan Yuk}
\affil{Homer L.\ Dodge Department of Physics and Astronomy,
University of Oklahoma, Norman, OK 73019, USA}
\email{hyuk@ou.edu}

\author[0000-0001-9203-2808]{Xinyu Dai}
\affil{Homer L.\ Dodge Department of Physics and Astronomy,
University of Oklahoma, Norman, OK 73019, USA}
\email{xdai@ou.edu}

\author{T. Jayasinghe}
\affiliation{Department of Astronomy, The Ohio State University, 140 West 18th Avenue, Columbus, OH 43210, USA}

\author[0000-0001-9608-6395]{Hai Fu}
\affil{Department of Physics \& Astronomy, University of Iowa, Iowa City, IA 52242, USA}

\author[0000-0002-6821-5927]{Hora D. Mishra}
\affil{Homer L.\ Dodge Department of Physics and Astronomy,
University of Oklahoma, Norman, OK 73019, USA}

\author[0000-0001-6017-2961]{Christopher~S.~Kochanek}
\affiliation{Department of Astronomy, The Ohio State University, 140 West 18th Avenue, Columbus, OH 43210, USA}
\affiliation{Center for Cosmology and AstroParticle Physics, The Ohio State University, 191 W.\ Woodruff Ave., Columbus, OH 43210, USA}

\author[0000-0003-4631-1149]{Benjamin J. Shappee}
\affiliation{Institute for Astronomy, University of Hawai\`{}i at Manoa, 2680 Woodlawn Dr., Honolulu, HI 96822, USA}

\author{K.~Z.~Stanek}
\affiliation{Department of Astronomy, The Ohio State University, 140 West 18th Avenue, Columbus, OH 43210, USA}
\affiliation{Center for Cosmology and AstroParticle Physics, The Ohio State University, 191 W.\ Woodruff Ave., Columbus, OH 43210, USA}

\begin{abstract}

Low luminosity active galactic nuclei (LLAGN) probe accretion physics in the low Eddington regime and can provide additional clues about galaxy evolution. AGN variability is ubiquitous and thus provides a reliable tool for finding AGN. We analyze the All-Sky Automated Survey for SuperNovae light curves of 1218 galaxies with $g<14$  mag and Sloan Digital Sky Survey spectra in search of AGN. We find 37 objects that are both variable and have AGN-like structure functions, which is about 3\% of the sample. The majority of the variability selected AGN are LLAGN with Eddington ratios ranging from $10^{-4}$ to $10^{-2}$. We thus estimate the fraction of LLAGN in the population of galaxies as 2\% down to a median Eddington ratio of $2\times10^{-3}$. Combining the BPT line ratio diagnostics and the broad-line AGN, up to $\sim$60\% of the AGN candidates are confirmed spectroscopically. The BPT diagnostics also classified 10--30\% of the candidates as star forming galaxies rather than AGN.

\end{abstract}

\section{Introduction}
Active galactic nuclei (AGNs) are some of the most luminous objects found in the universe, driven by accretion onto a supermassive black hole. Studies show that AGN and their host galaxy properties are strongly correlated. For example, the AGN class is closely related to galaxy morphology (e.g., \citealt{slavchevamihova11}; \citealt{chen17}; \citealt{bornancini18}; \citealt{gkini21}), AGN feedback can suppress star formation (e.g., \citealt{hopkins10}; \citealt{zubovas12}; \citealt{alberts16}; \citealt{yuan18}), and merger events can ignite AGN activities (e.g., \citealt{keel85}; \citealt{ellison11}, \citeyear{ellison19}; \citealt{koss12}). The strongest evidence of coevolution between AGN and host galaxies is the $M-\sigma$ correlation between black hole masses and the stellar velocity dispersion (\citealt{ferrarese00}; \citealt{gebhardt00}; \citealt{graham11}; \citealt{mcconnell11}; \citealt{morabito12}). The central black hole is active only some of the time, so the fraction of active galaxies constrains the AGN duty cycle, which in turn can be used to match the black hole with accretion rate history (e.g., \citealt{yu02}; \citealt{marconi04}; \citealt{tamura06}; \citealt{shankar09}). AGN activities also have an environmental dependence: rich clusters show lower AGN activity due to inefficient mergers (e.g., \citealt{kauffmann04}; \citealt{lopes17}; \citealt{mishra20}) at low redshifts below the peak of AGN activities. At high redshifts, observational evidence points to higher AGN fractions in clusters (\citealt{eastman07}; \citealt{haggard10}; \citealt{martini13}). For the low-density environments in cosmic voids, the existing few studies suggest potential higher AGN fractions (\citealt{constantin08}; \citealt{mishra21}).

An interesting subset of AGN is the low luminosity AGN (LLAGN). There exist many, some fundamental, differences between LLAGN and luminous AGN. LLAGN are thought to be less triggered by major mergers and more by effects such as tidal disruptions and disk instabilities (e.g., \citealt{hopkins09}; \citealt{hopkins14}). Observations and models suggest that the broad-line region and the obscuring ``torus" disappear for the AGN with the lowest luminosities (e.g., \citealt{ho97apjs2}; \citealt{nicastro00}; \citealt{laor03}; \citealt{elitzur06}), so LLAGN do not follow the standard AGN unification model. Instead, the black hole may have a radiatively inefficient accretion flow at small radii and a truncated thin disk farther out combined with a jet or outflow (e.g., \citealt{chen89}; \citealt{ho08}). AGN with lower luminosities tend to have a greater fraction of their accretion power converted into a relativistic jet which can have a significant impact on galaxy formation \citep{ho08}. Active galaxies are known to spend significantly more time as LLAGN than as luminous AGN, but LLAGN evolution is not well understood. One observation that can provide clues about the evolution of AGN is the LLAGN fraction as a constraint on the duty cycle of this form of accretion.

One of defining features of AGN is their variability across the entire electromagnetic spectrum on all timescales. The mechanism responsible for the variability is still unknown. Proposed mechanisms include thermal fluctuations (e.g., \citealt{lightman74}; \citealt{shakura76}; \citealt{kelly09}), accretion disk instabilities (e.g., \citealt{kawaguchi98}; \citealt{trevese02}), supernova explosions (e.g., \citealt{aretxaga97}), and reprocessing photons between the accretion disk and the corona (e.g., \citealt{haardt91}; \citealt{mchardy18}). It is known that the variability is stochastic in nature (e.g., \citealt{kelly09}; \citealt{macleod10}) and it can be modeled by simple stochastic process models.


As a ubiquitous feature of AGN, variability can be used to find AGN. One way of characterizing the AGN variability is the structure function (e.g., \citealt{vandenberk04}; \citealt{kozlowski17}; \citealt{Middei_2017}), defined by the mean square flux change as a function of the time separating the measurements. The advantage of the structure function is that it can be measured accurately even with unevenly sampled light curves with seasonal gaps. The structure functions of AGN are roughly a broken power law, initially rising in amplitude as the time lag increases and flattening into a plateau (e.g., \citealt{trevese94}; \citealt{macleod10}; \citealt{kozlowski17}). The light curve of any galaxy can be examined to see if it has an AGN-like structure function to identify AGN that other methods, such as broad emission lines, optical color space, X-ray, mid-IR, and radio emissions, may miss.

There have been previous studies of variability selected AGN samples. \citet{macleod11} analyzed $\sim$10,000 variable sources from Sloan Digital Sky Survey (SDSS) data. They compared the structure function of the variable objects with the damped random walk (DRW) model to select quasars, and their fraction of variability selected quasars that were previously confirmed quasars was $>$90\%. \citet{butler11} parameterized the structure function of quasars as a function of brightness and used this model to find AGN. Their algorithm identified $99\%$ of the known quasars correctly while finding thousands of new quasars, increasing the quasar sample size by $29\%$. \citet{ruan12} investigated the optical variability of $\sim$190,000 variable objects from the MIT Lincoln Laboratory (MITLL) Lincoln Near-Earth Asteroid Research (LINEAR) survey. They characterized the blazar variability by the DRW model and compared it to the LINEAR variables. Compared to the Fermi catalog, their selection method identified blazars with an efficiency of $E\geq88\%$ and a completeness of $C=88\%$. \citet{sanchezsaez19} analyzed the variability features of over 200,000 light curves from QUEST-La Silla AGN Survey, finding $\sim5000$ AGN. Their spectroscopic follow-up found that $>$80\% of their variability selected AGN display AGN-like spectral features.

Most of the variability selected AGN samples are quasars at the high luminosity end of the AGN population. In this study, we seek to find AGN with a wider range of luminosity, including LLAGN, so we can place tighter constraints on the AGN fraction among galaxies in the low Eddington regime. We use the data from the All-Sky Automated Survey for SuperNovae (ASAS-SN; \citealt{shappee14}; \citealt{asassn2017}) to do a variability analysis of 1218 nearby SDSS galaxies using structure functions to identify AGN candidates. The SDSS spectra \citep{sdss16} can then be used to further characterize the candidates.

In Section \ref{secdata}, we describe the ASAS-SN data used for this study. In Section \ref{secmethod}, we discuss the systematic procedures for analyzing the light curves to find AGN candidates. In Section \ref{secresults}, we present our results. In Section \ref{secanalysis}, we further analyze the properties of the candidates. The paper is summarized in Section \ref{secconclusion}. Throughout this paper, we assume a $\rm{\Lambda}$CDM cosmological model, with $H_0=70 \textrm{ km s}^{-1} \textrm{ Mpc}^{-1}$, $\Omega_{\Lambda}=0.7$, and $\Omega_M=0.3$.

\section{Data}
\label{secdata}
ASAS-SN (\citealt{shappee14}; \citealt{asassn2017}) is a network of telescopes that has been observing the extragalactic sky since 2012. ASAS-SN has expanded a number of times and now consists of 20 14 cm telescopes in five mounts that have been surveying the entire visible sky every night since late 2017. The original two mounts used the V-band filter until 2018 and then switched to the g-band, while the three new mounts commissioned in late 2017 used the g-band from the start. ASAS-SN’s limiting magnitude is $V \sim 16.5-17.5$ and $g \sim 17.5-18.5$ depending on lunation. The field of view of an ASAS-SN camera is 4.5 deg$^2$, the pixel scale is 8\farcs0 and the FWHM is typically ${\sim}2$ pixels. The ASAS-SN light curves were extracted as described in \citet{2018MNRAS.477.3145J} using image subtraction \citep{1998ApJ...503..325A,2000A&AS..144..363A} and aperture photometry on the subtracted images with a 2 pixel radius aperture. The APASS catalog \citep{2015AAS...22533616H} was used for calibration. We corrected the zero-point offsets between the different cameras as described in \citet{2018MNRAS.477.3145J}. The photometric errors were recalculated as described in \citet{2019MNRAS.485..961J}.

For this study, we extracted nuclear light curves for the 1218 $g<14$ mag SDSS galaxies with SDSS spectra \citep{sdss16}. The spectrum allows us to spectroscopically classify the variability identified candidates. A typical object has 280 V-band data points spanning over 2000 days, and 180 g-band data points spanning over 600 days. The typical uncertainties are 0.02 mag for the V-band and 0.03 mag for the g-band. Before starting the analysis of the light curves, significant outliers are removed, as they can affect the variability measurements. We discarded data points more than $5\sigma$ from the median, where $\sigma$ is the standard deviation of the measurements. We also rejected objects with mean ASAS-SN magnitudes that differ from the SDSS magnitudes by $>$2 mag; most of these objects have nearby bright sources that can contaminate the photometry. The ASAS-SN photometry data is given in Table \ref{LCtable}.

\begin{deluxetable}{ccccc}
\tablecaption{ASAS-SN photometry of AGN candidates}
\tablecolumns{5}
\tablehead{\colhead{Name} & \colhead{Filter} & \colhead{Date (HJD)} & \colhead{Magnitude} & \colhead{Mag err}}
\startdata
NGC 0863  & V   & 2456615.77671 & 13.720 & 0.026 \\
          & V   & 2456618.77137 & 13.738 & 0.012 \\
          & V   & 2456631.84639 & 13.708 & 0.015 \\
          &     & ... & ... & ... \\
UGC 02018 & V   & 2456220.99995 & 13.887 & 0.010 \\
          & V   & 2456227.92043 & 13.904 & 0.023 \\
          & V   & 2456308.79893 & 13.900 & 0.018 \\
          &     & ... & ... & ... \\
NGC 0988  & V   & 2456608.89271 & 13.808 & 0.042 \\
          & V   & 2456630.83460 & 13.712 & 0.030 \\
          & V   & 2456636.81180 & 13.464 & 0.015 \\
          &     & ... & ... & ... \\
NGC 0988  & g   & 2458013.77572 & 13.809 & 0.023 \\
          & g   & 2458014.77038 & 13.806 & 0.023 \\
          & g   & 2458015.76754 & 13.806 & 0.023 \\
          &     & ... & ... & ... \\
\hline
\enddata
\tablecomments{This table is available in its entirety in the machine-readable tables.}
\label{LCtable}
\end{deluxetable}

\section{Method}
\label{secmethod}
We searched for AGN in two steps. First, we calculate the excess variance of each light curve to see if the source has significant variability. If a source appears to be variable, we calculate the structure function of the light curve, and model it to see if the variability is AGN-like. We analyzed the V- and g-band light curves separately for each target.

\subsection{Excess variance}
The excess variance measures the intrinsic variability of the object beyond measurement uncertainties (e.g., \citealt{nandra97}; \citealt{edelson02}; \citealt{Vaughan_2003}). For this study, we calculate the excess variance using magnitudes, so we use unnormalized excess variance, which is calculated by correcting the variance for the photometric noise,

\begin{equation}
    \sigma_{\mathrm{XS}}^2 = S^2-\overline{\sigma_{\mathrm{err}}^2},
    \label{EV_eq}
\end{equation}
where $S^2$ is the variance of the light curve and $\overline{\sigma_{\mathrm{err}}^2}$ is the mean square measurement error. If this value is negative, then the variance of the data is within the range of measurement error. Assuming the uncertainty of $\overline{\sigma_{\mathrm{err}}^2}$ is negligible, the error in the excess variance is

\begin{equation}
    \Delta(\sigma_{\mathrm{XS}}^2) = \sqrt{\frac{2}{N-1}}S^2,
\end{equation}
where $N$ is the number of data points (\citealt{trumpler53}; \citealt{edelson02}).

\begin{figure}
\centering
\includegraphics[width=6.5in]{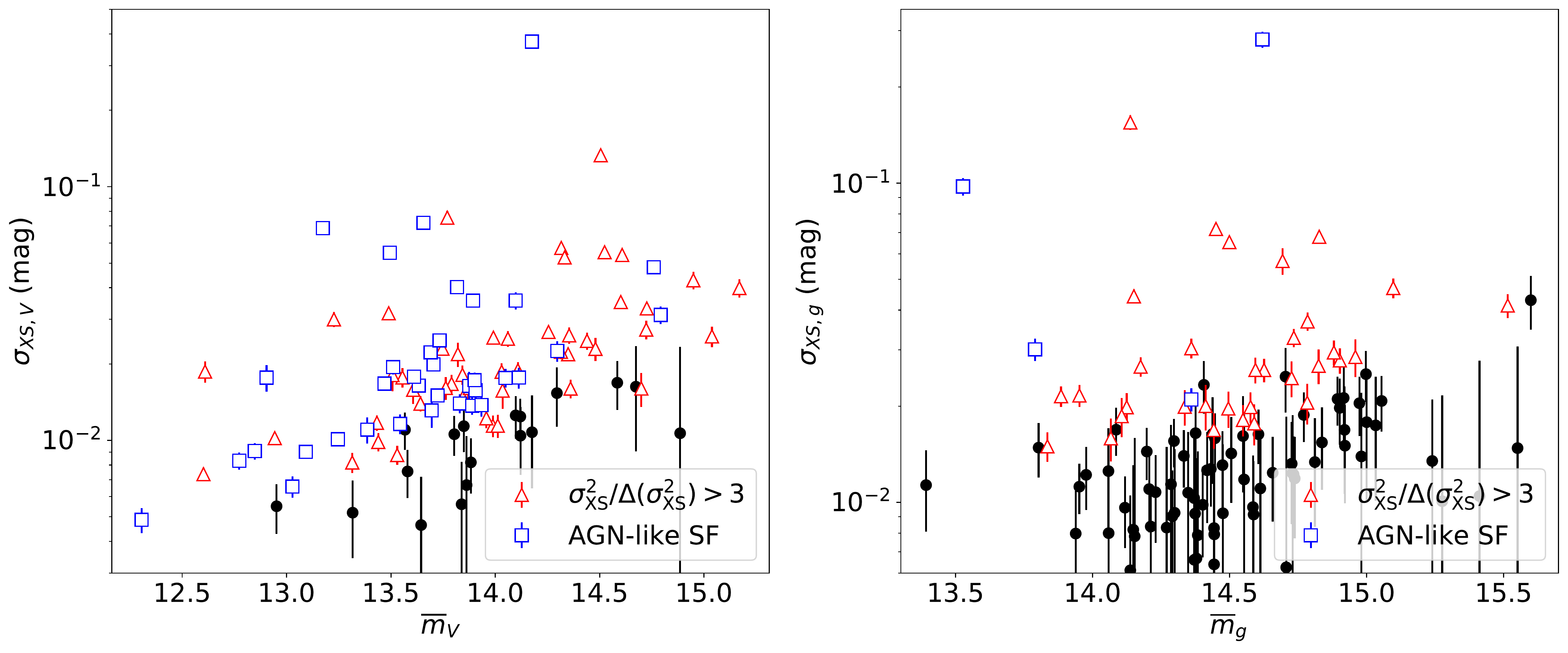}
\caption{Excess variance versus mean magnitude of all ASAS-SN objects with $\sigma_{\mathrm{XS}}^2>0$. The black circles are galaxies with $\sigma_{\mathrm{XS}}^2 / \Delta(\sigma_{\mathrm{XS}}^2) < 3$, red triangles are galaxies with $\sigma_{\mathrm{XS}}^2 / \Delta(\sigma_{\mathrm{XS}}^2) > 3$, and the blue squares also have AGN-like structure functions.}
\label{EV_plot}
\end{figure}

\begin{figure}
\centering
\includegraphics[width=4in]{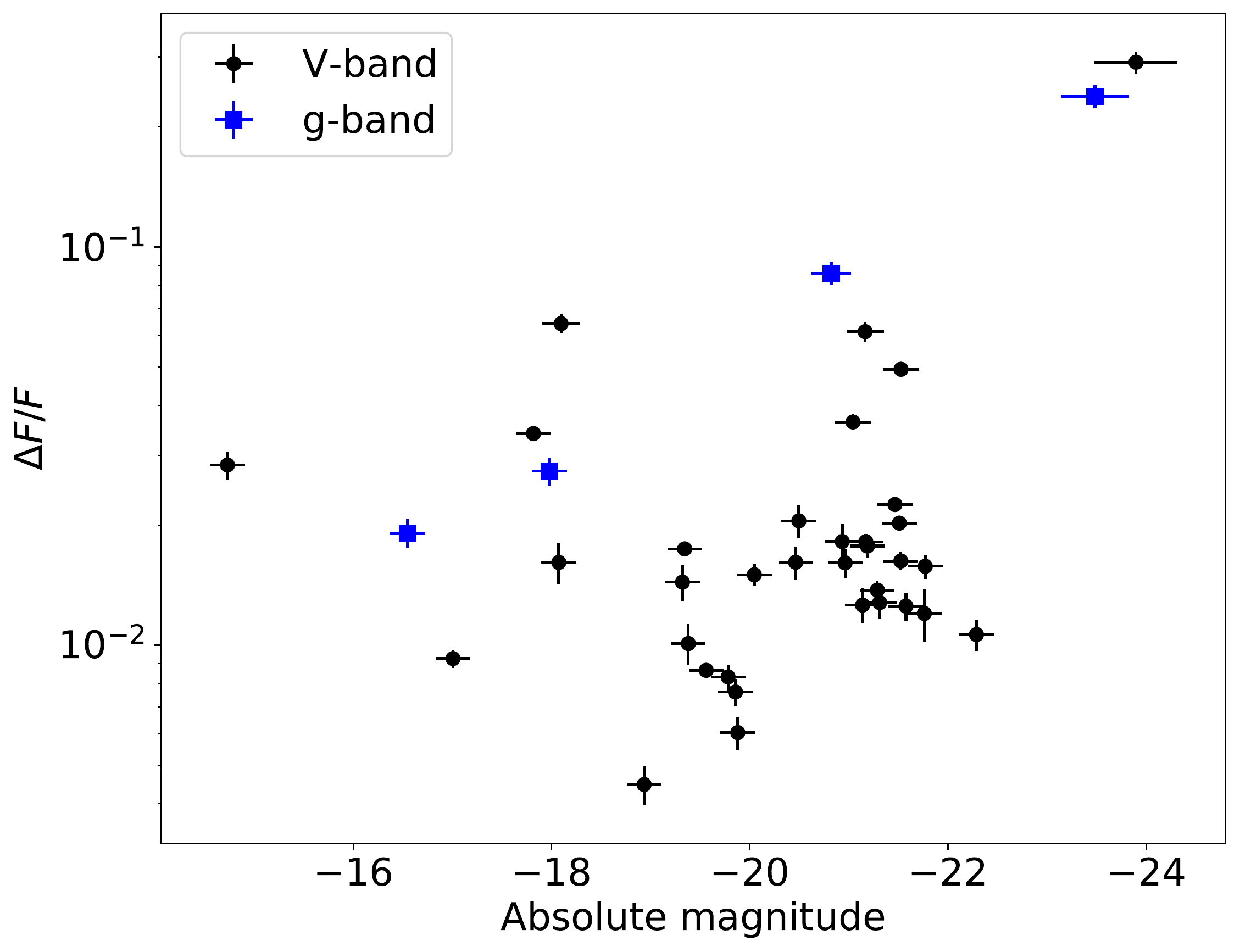}
\caption{Fractional flux variability versus absolute magnitude of AGN candidates. Three objects are classified as AGN candidates in both V- and g-band, and for these objects, both V- and g-band variabilities are plotted. The variability amplitude of these objects agrees to within a factor of 3.}
\label{var_mag_plot}
\end{figure}

The distribution of the galaxies in $\sigma_{\mathrm{XS}}$ is shown in Figure \ref{EV_plot}, and a galaxy is considered to have significant variability if $\sigma_{\mathrm{XS}}^2 / \Delta(\sigma_{\mathrm{XS}}^2) > 3$. Using the variability in magnitudes, the fractional flux variability is

\begin{equation}
    \frac{\Delta F}{F} = |1 - 10^{-0.4 \sigma_{\mathrm{XS}}}|,
\end{equation}
and as seen in Figure \ref{var_mag_plot}, most objects with $\sigma_{\textrm{XS}}^2/\Delta \sigma_{\textrm{XS}}^2>3$ have $\Delta F/F\gtrsim10^{-2}$. Excluding highly variable outliers, there is a weak trend for fainter objects to require higher variability amplitudes for detection, which is to be expected given the increase in photometric uncertainties with magnitude.

\subsection{Structure function}

Several different definitions of the structure function are in use, where the two most common forms are

\begin{equation}
    \textrm{SF}(\tau) = \sqrt{\langle[m(t+\tau) - m(t)]^2 \rangle - \sigma_{noise}^2},
    \label{SF_eq}
\end{equation}
(e.g., \citealt{bauer09}; \citealt{Middei_2017}) and

\begin{equation}
    \textrm{SF}(\tau) = \sqrt{\frac{\pi}{2}\langle|m(t+\tau) - m(t)| \rangle^2 - \sigma_{noise}^2},
    \label{SF_alt_eq}
\end{equation}
(e.g., \citealt{diclemente96}; \citealt{vandenberk04}), where $m(t)$ is the magnitude at time $t$, $\tau$ is the time difference between the data pair, and

\begin{equation}
    \sigma_{noise}^2 = \langle \sigma_{\mathrm{err}}^2(t) + \sigma_{\mathrm{err}}^2(t+\tau) \rangle
    \label{sig_noise_eq}
\end{equation}
is the noise contribution where $\sigma_{\mathrm{err}}(t)$ is the uncertainty in the measured magnitude. For this study, we use Equation \ref{SF_eq}. The structure function is calculated in bins of width $\Delta \log(\tau/\rm{day})=0.1$. The uncertainty of the structure function is

\begin{equation}
    \Delta \textrm{SF} = \frac{1}{N_{\mathrm{bin}}\cdot \textrm{SF}}
    \sqrt{\sum_{t,\tau} \bigg\{ [m(t+\tau) - m(t)]^2
    \cdot [\sigma_{\mathrm{err}}^2(t+\tau) + \sigma_{\mathrm{err}}^2(t)]\bigg\}},
    \label{SF_err_eq}
\end{equation}
where $N_{\mathrm{bin}}$ is the number of data point pairs in the bin and we are assuming that the uncertainty in $\sigma_{\mathrm{err}}$ is negligible. For small $\tau$, the errors are large, so we discard time scales $\tau<3$ days.

We next model the structure function as a power law plus a constant, $\textrm{SF}(\tau)=A \tau^B + C$, representing the contributions from the AGN plus a white noise contribution from the measurement error. We fit the structure function both with an AGN ($A\neq 0$) and just as white noise ($A=0$), and then use an F-test to determine if the fit with the power-law component is significantly better than the constant fit. The distribution of F-test results is bimodal, with peaks at around 0.02 and 0.40. To filter out the clearly bad candidates while still keeping some borderline candidates, we keep candidates with F-test values $<0.1$.

In addition to F-test, we examined the power-law index, $B$, to determine if the structure function is AGN-like. 
Quasar variability is a form of red noise, in the sense that the amplitude of the variability initially increases with time scale, but then seems to saturate on longer time scales (years to decades, e.g., \citealt{macleod12}; \citealt{kozlowski17}). If the DRW was truly the correct model, $B=0.5$ on short time scales and then a power-law fit will become shallower ($B<0.5$) on time scales similar to the damping time scale or longer. \citet{schmidt10} fit power-law structure functions to the SDSS Stripe 82 quasars and used this to set a selection limit of $B>0.055$. The Stripe 82 light curves are moderately longer (8 yr) than either the ASAS-SN V- or g-band light curves and poorly sampled on shorter time scales, so we would expect the typical $B$ found here to be larger. We find values that are relatively steep (median $B=1.5$), but we suspect the exact values are both noisy and affected by systematic errors.  We found that an empirical selection cut of $B>0.3$ appears to work well.
Then we examined higher slopes in detail to determine the upper limit. Structure functions with $2<B<3$ tend to display an upward trend starting at $\log(\tau\textrm{/day})>2.5$, while the structure functions with smaller slopes start showing such trend earlier. Structure functions with $3<B<4$ display the general upward trend, however, the shape is more complicated where the simple power law plus a constant model does not fit well. For $B>4$, the structure function is mostly flat, with only the last few data points slightly larger than the others, so we set the upper limit of $B$ for the AGN candidate at 4 and a reliable slope measure at 3. Before the final confirmation of a candidate, the light curves and SF are visually inspected.

\begin{figure}
\centering
\includegraphics[width=7in]{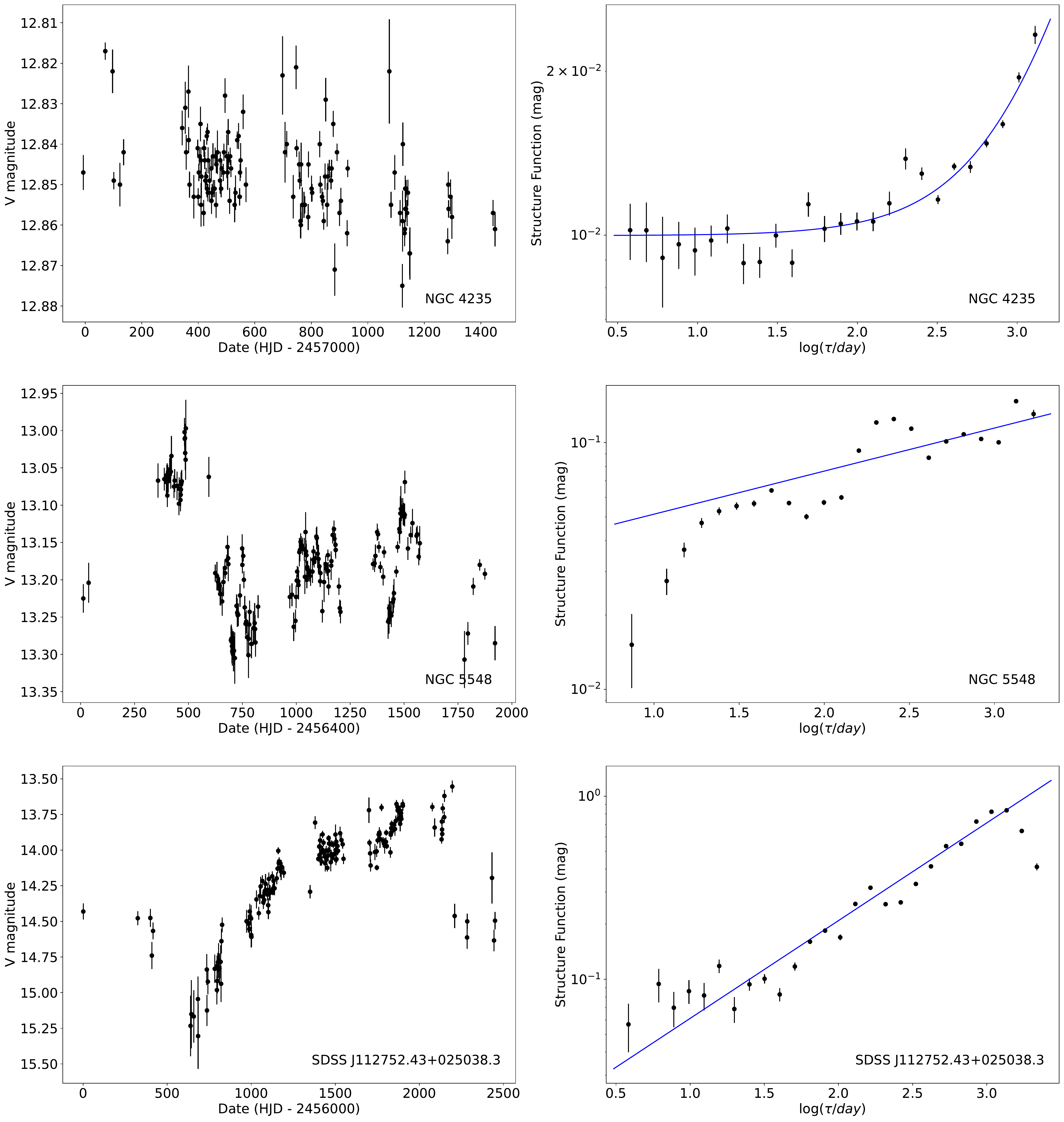}
\caption{The ASAS-SN V-band light curves and the structure functions of NGC 4235 and NGC 5548, two known AGN, and a new variability selected AGN candidate, SDSS J112752.43+025038.3. The blue lines are the best fit structure function models.}
\label{SF_plot}
\end{figure}

In addition, high variability objects ($\sigma_{\textrm{XS}}>0.03$) that fail the SF criteria are also manually examined. Some candidates are lost because they have more complex structure functions, such as the broken power law. Once an object passes all of these criteria in either V- or g-band, it is considered an AGN candidate. Figure \ref{SF_plot} shows the light curve and structure function of the three candidates. One of the structure functions has a small peak at $\log(\tau/\textrm{day})\sim2.2$, which corresponds to a time lag of $\sim150$ days. This is a common feature observed in many of the structure functions and it is probably an artifact of the seasonal gaps.

\begin{figure}
    \centering
    \includegraphics[width=5.5in]{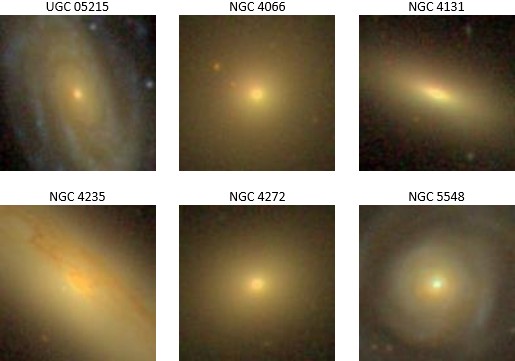}
    \caption{The SDSS images \citep{sdss16} of the 6 of the 37 AGN candidates. Each image is 50$^{\prime\prime}$ $\times$ 50$^{\prime\prime}$ with North up and East left.}
    \label{sdss_cutout}
\end{figure}

\section{Results}
\label{secresults}

Of the 1218 galaxies, 88 objects in the V-band and 38 objects in the g-band (10 objects in both) display significant variability ($\sigma_{\mathrm{XS}}^2 / \Delta(\sigma_{\mathrm{XS}}^2) > 3$). Of these, 36 objects in the V-band and 4 objects in the g-band pass the structure function test. With three of them overlapping in both bands, we find a total of 37 AGN candidates, eight of which were previously classified as AGN by SDSS \citep{sdss16}. Table \ref{AGNtable} summarizes the candidates, and Figure \ref{sdss_cutout} shows the images of six.

\begin{longrotatetable}
\begin{deluxetable}{lclclcc}
\tablecaption{Data of AGN candidates from ASAS-SN light curves}
\tablecolumns{6}
\tablehead{\colhead{Object name} & \colhead{Known AGN} & \colhead{Redshift} & \colhead{Morphology} & \colhead{$\log L/L_{\odot}$} & \colhead{$\log M_{\rm{BH}}/M_{\odot}$} & \colhead{Eddington Ratio}}
\startdata
NGC 0863 & Yes & 0.0261 & SA(s)a: & 10.63 & 8.0 & 1.0$\times10^{-3}$ \\
UGC 02018 &  & 0.0206 & SB0$^+$(rs): & 10.30 & 8.3 & 7.4$\times10^{-4}$ \\
NGC 0988 &  & 0.0050 & SB(s)cd: & \hspace{5pt}9.16 & 7.1 & 3.4$\times10^{-3}$ \\
NGC 2552 &  & 0.0018 & SA(s)m? & \hspace{5pt}7.82 & 5.7 & 1.6$\times10^{-3}$ \\
UGC 05215 &  & 0.0183 & Sbc & 10.11 & 8.1 & 8.1$\times10^{-4}$ \\
NGC 3102 &  & 0.0101 & S0$^-$: & \hspace{5pt}9.65 & 7.6 & 7.3$\times10^{-4}$ \\
NGC 3499 &  & 0.0051 & I0? & \hspace{5pt}9.05 & 7.0 & 1.7$\times10^{-3}$ \\
NGC 3594 &  & 0.0209 & SB0: & 10.12 & 8.1 & 1.0$\times10^{-3}$ \\
SDSS J112752.43+025038.3 &  & 0.0897 & Unknown & 11.48 & 9.5 & 1.4$\times10^{-2}$ \\
NGC 3835 &  & 0.0081 & Sab: edge-on & \hspace{5pt}9.68 & 7.6 & 5.1$\times10^{-4}$ \\
NGC 3886 &  & 0.0194 & S0$^-$: & 10.40 & 8.4 & 8.7$\times10^{-4}$ \\
NGC 3978 & Yes & 0.0332 & SABbc: & 10.84 & 8.8 & 5.1$\times10^{-4}$ \\
NGC 4041 &  & 0.0040 & SA(rs)bc: & \hspace{5pt}9.50 & 7.5 & 2.3$\times10^{-4}$ \\
NGC 4043 &  & 0.0214 & (R)SB(r)0$^0$: & 10.39 & 8.4 & 9.0$\times10^{-4}$ \\
NGC 4061 &  & 0.0245 & E: & 10.53 & 8.9 & 4.7$\times10^{-4}$ \\
NGC 4062 &  & 0.0025 & SA(s)c & \hspace{5pt}8.73 & 6.7 & 4.9$\times10^{-4}$ \\
NGC 4066 &  & 0.0245 & E & 10.51 & 8.8 & 5.2$\times10^{-4}$ \\
NGC 4070 & Yes & 0.0239 & E & 10.53 & 8.9 & 3.8$\times10^{-4}$ \\
NGC 4080 &  & 0.0018 & Im? & \hspace{5pt}8.10 & 6.0 & 1.8$\times10^{-3}$ \\
NGC 4092 &  & 0.0225 & S? & 10.31 & 8.3 & 8.0$\times10^{-4}$ \\
NGC 4095 &  & 0.0238 & E? & 10.45 & 8.8 & 3.0$\times10^{-4}$ \\
NGC 4131 &  & 0.0124 & S & \hspace{5pt}9.94 & 7.9 & 7.6$\times10^{-4}$ \\
NGC 4158 & Yes & 0.0082 & SA(r)b: & \hspace{5pt}9.66 & 7.6 & 7.8$\times10^{-4}$ \\
NGC 4213 &  & 0.0224 & E & 10.44 & 8.8 & 3.2$\times10^{-4}$ \\
NGC 4224 &  & 0.0086 & SA(s)a: edge-on & \hspace{5pt}9.88 & 7.8 & 3.0$\times10^{-4}$ \\
NGC 4233 & Yes & 0.0076 & S0$^0$ & \hspace{5pt}9.87 & 7.8 & 3.9$\times10^{-4}$ \\
NGC 4235 & Yes & 0.0075 & SA(s)a edge-on & \hspace{5pt}9.84 & 7.2 & 7.3$\times10^{-4}$ \\
NGC 4241 &  & 0.0025 & SB(s)cd & \hspace{5pt}8.09 & 6.0 & 2.4$\times10^{-3}$ \\
NGC 4272 &  & 0.0282 & E & 10.55 & 8.9 & 2.9$\times10^{-4}$ \\
NGC 4944 &  & 0.0232 & S0/a? edge-on & 10.53 & 8.5 & 2.4$\times10^{-3}$ \\
NGC 5032 &  & 0.0213 & SB(r)b & 10.34 & 8.3 & 1.8$\times10^{-3}$ \\
NGC 5162 &  & 0.0227 & Scd: & 10.38 & 8.4 & 6.2$\times10^{-4}$ \\
UGC 08516 &  & 0.0034 & Scd: & \hspace{5pt}8.66$^a$ & 6.5 & 1.0$\times10^{-3}$ \\
NGC 5273 & Yes & 0.0036 & SA0$^0$(s) & \hspace{5pt}9.16 & 6.5 & 1.5$\times10^{-3}$ \\
IC 4345 &  & 0.0311 & E? & 10.63 & 9.0 & 3.6$\times10^{-4}$ \\
NGC 5548 & Yes & 0.0163 & (R')SA(s)0/a & 10.39 & 7.8 & 5.2$\times10^{-3}$ \\
NGC 5633 &  & 0.0077 & (R)SA(rs)b & \hspace{5pt}9.75 & 7.7 & 4.2$\times10^{-4}$ \\
\hline
\enddata
\tablecomments{The objects are listed in ascending order of R.A. The redshift and morphology are from SDSS \citep{sdss16} and the NASA/IPAC Extragalactic Database (NED) \citep{NED}, respectively. The luminosity refers to the object's V-band (or g-band, if noted) luminosity. The central supermassive black hole mass and the Eddington ratio are estimates, as described in section \ref{estimate}.\\
$^a$ g-band luminosity.}
\label{AGNtable}
\end{deluxetable}
\end{longrotatetable}

\section{Analysis}
\label{secanalysis}

\subsection{Comparison with spectral classification}

\begin{figure}
\centering
\includegraphics[width=6.5in]{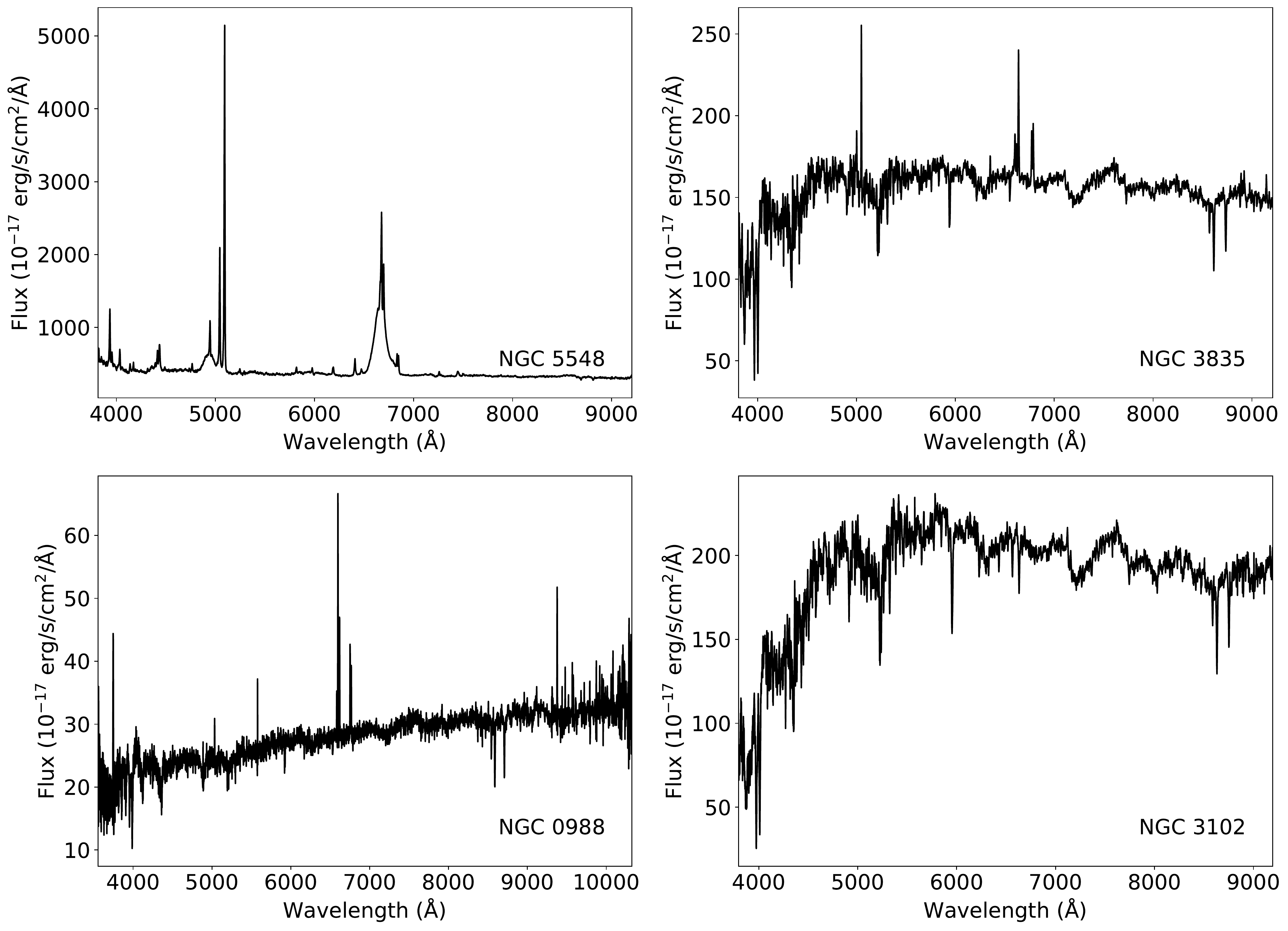}
\caption{SDSS Spectra of several AGN candidates. Top left: a spectrum of the known AGN NGC 5548 showing a clear broad H$\alpha$ emission line. Top right: a spectrum with a narrow H$\alpha$ emission line that is classified as an AGN based on its BPT line ratios. Bottom left: a spectrum with a narrow H$\alpha$ emission line that is not classified as an AGN given its line ratios. Bottom right: a spectrum with weak or absent H$\alpha$ emission. There are 4, 18, 7, and 8 candidates in each of these categories.}
\label{spectra}
\end{figure}

Since all the targets have SDSS spectra, we can examine and measure several spectral features to further understand the properties of our variability selected AGN sample. Examples of the spectra are shown in Figure \ref{spectra}. Only four of the spectra (NGC 0863, NGC 4235, NGC 5273, and NGC 5548) display clear broad H$\alpha$ emission lines, and all four are previously known AGN. We used the IDL package \texttt{SPFIT} \citep{fu18} to fit the spectra. \texttt{SPFIT} simultaneously fits the stellar continuum and emission lines, using the Penalized Pixel-Fitting method (pPXF; \citealt{cappellari04}) in three steps. First, the spectral regions with possible emission lines are masked to fit the stellar continuum. Next, the best-fit continuum is subtracted and the remaining emission-line-only spectrum is fit. Finally, the nonlinear optimizer package \texttt{MPFIT} \citep{markwardt09} uses these two models as the initial values for a fit to the full spectrum. 

\begin{figure}
\centering
\includegraphics[width=6.5in]{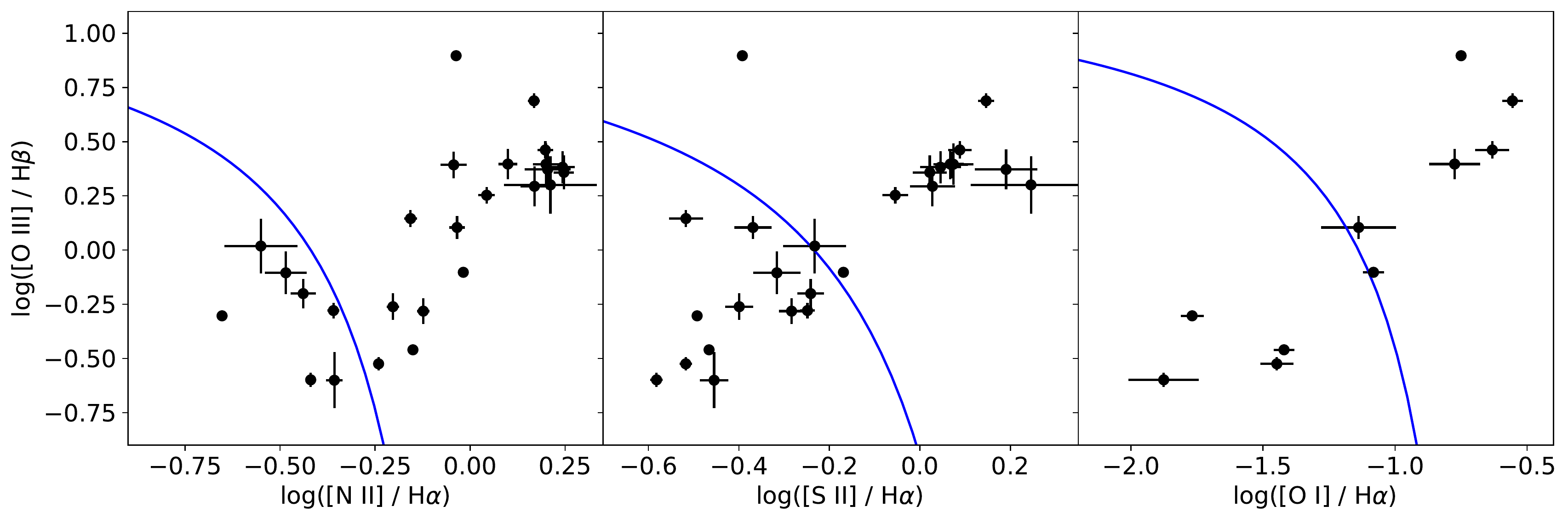}
\caption{BPT diagrams of the AGN candidates. The blue curve is the \citet{Kewley_06} classification line dividing star forming sources (lower right) from AGN (upper right).}
\label{BPT_plot}
\end{figure}

We then use Baldwin--Phillips--Terlevich (BPT) diagrams \citep{Baldwin_81} to classify and confirm the AGN candidates with emission-line flux measurements with signal-to-ratios $>3$ (Figure \ref{BPT_plot}). Because \texttt{SPFIT} is not designed to fit spectra with broad emission lines, three of four spectra with broad H$\alpha$ emission are not well fit and are excluded from the BPT analysis. Figure \ref{BPT_plot} shows the BPT diagrams of the AGN candidates with the star forming/AGN dividing line from \citet{Kewley_06}. In all three panels, a large fraction of the AGN candidates are confirmed based on the BPT diagnostics: 19 out of 26 for $\log( \textrm{\OIII/\Hb} )$ vs.\ $\log(\textrm{\NII/\Ha})$, 14 out of 26 for $\log( \textrm{\OIII/\Hb} )$ vs.\ $\log(\textrm{\SII/\Ha})$, and 6 out of 10 for $\log( \textrm{\OIII/\Hb} )$ vs.\ $\log(\textrm{\OI/\Ha})$.
For the most commonly used $\log( \textrm{\OIII/\Hb} )$ vs.\ $\log(\textrm{\NII/\Ha})$ diagnostic, the AGN confirmation rate is 73\%. When including the three other broad line AGN, the AGN confirmation rate is 85\%. Out of the total sample of 37 candidates, the confirmation rate is 60\%. There exist prior emission line measurements and BPT classifications from SDSS spectra by MPA-JHU \citep{brinchmann04}, which give classifications largely consistent with our results.

However, a significant fraction of AGN candidates are classified as star-forming galaxies and it is possible that some of the galaxies in the AGN region of the BPT diagrams are AGN imposters that are actually retired galaxies \citep{cidfernandes11}. Such galaxies can be identified by the equivalent width of their H$\alpha$ emission line; galaxies with $\textrm{EW}_{\textrm{H}\alpha}<3$ \AA\hspace{1pt} are considered retired galaxies. If we also require $\textrm{EW}_{\textrm{H}\alpha}>3$ \AA\hspace{1pt}, the AGN fraction based on the BPT diagrams drops down to 20--40\%, suggesting that the emission line diagnostic has its own limitations.

\subsection{Low-luminosity AGN}

\begin{figure}
\centering
\includegraphics[width=6.5in]{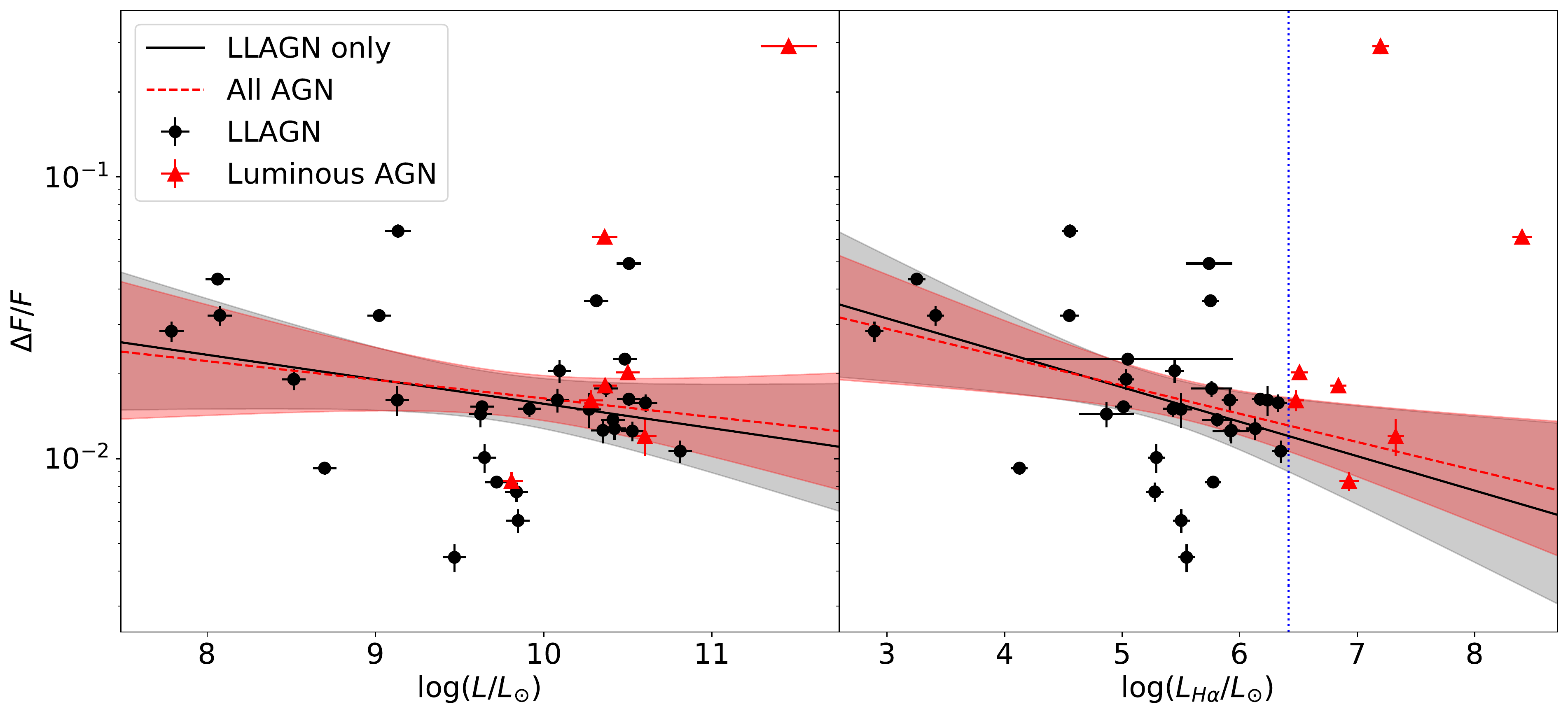}
\caption{Fractional flux variability of the candidates versus host V/g-band (left) and H$\alpha$ (right) luminosities with the best linear fits (shaded areas represent the 90\% confidence regions).  The dotted blue vertical line on the right graph is the dividing line for LLAGN ($L_{\textrm{H}\alpha}=10^{40}$ erg s$^{-1}$). The outlier at $\Delta F/F\sim0.3$ is excluded for both fits, and an additional outlier at $\log(L_{\textrm{H}\alpha}/L_{\odot})\sim8.5$ is excluded for the $\Delta F/F$ versus $\log(L_{\textrm{H}\alpha}/L_{\odot})$ fit.}
\label{var_lum_fit_plot}
\end{figure}

Objects with $L_{\textrm{H}\alpha}<10^{40}$ erg s$^{-1}$ are considered LLAGN \citep{ho97apjs1}. After manually estimating the H$\alpha$ luminosity of the broad-line AGN, 30 of 37 candidates--including NGC 5273, a galaxy with broad H$\alpha$ emission--are classified as LLAGN. Next, we took a closer look at the flux variability of the LLAGN compared to the host and the H$\alpha$ luminosities, as shown in Figure \ref{var_lum_fit_plot}. Galactic luminosities refer to the $V$- or $g$-band luminosities derived from the ASAS-SN photometry and are not bolometric. AGN variability is generally larger for less luminous AGN (e.g., \citealt{vandenberk04}; \citealt{macleod10}; \citealt{kozlowski16}). Broadly speaking, our candidates show the expected trends. We also observe that the slope is slightly steeper for the LLAGN, but the difference is not significant given the scatter. There is also a bias against detecting objects with smaller fractional variability.

\subsection{Black hole mass and Eddington luminosity}
\label{estimate}

We estimate the mass of the central supermassive black hole using the spheroid luminosity,

\begin{equation}
    \log \bigg(\frac{M_{\rm{BH}}}{M_{\odot}}\bigg) 
    = \alpha + \beta \log \bigg( \frac{L_{\textrm{sph}, I}}{10^{11} L_{\odot,I}} \bigg),
    \label{eq_mbh}
\end{equation}
where $L_{\rm{sph},I}$ is the $I$-band spheroid luminosity, and $\alpha=9.11\pm1.15$ and $\beta=1.02\pm0.11$ for quiescent galaxies and $\alpha=8.88\pm0.05$ and $\beta=1.02$ (fixed) for type 1 AGN \citep{Bennert_21}. We estimate $L_{\rm{sph},I}$ by the following procedure. We calculate the absolute magnitude using the mean magnitude in the $V$- or $g$-band measured by ASAS-SN, corrected for Galactic extinction \citep{Schlafly_11} and using \citet{Assef_10} for the K-correction. We also use the SED from \citet{Assef_10} to calculate the $V-I$ color and convert the $V$-band magnitudes into $I$-band magnitudes; $g$-band magnitudes are approximated as $V$-band magnitudes. Given the low resolution of ASAS-SN data, the flux is assumed to be the flux of the entire galaxy. For spiral galaxies, the flux is dominated by the disk. \citet{Bennert_21} show that the average difference between the spheroid and the disk magnitudes is $\langle m_{\textrm{sph}}-m_{\textrm{disk}} \rangle=0.795$ (Table 4 in \citealt{Bennert_21}). We use this correction for spiral galaxies and galaxies with unknown morphology. For elliptical galaxies, the entire galaxy is considered to be a spheroid. Then, using the Sun's absolute magnitude in the $I$-band \citep{Willmer_18}, each galaxy's $I$-band spheroid luminosity in terms of solar luminosity is obtained, which is then used to calculate the black hole mass.

The sources of uncertainties in $M_{\rm{BH}}$ calculation come from both ASAS-SN measurement uncertainties and systematic uncertainties. The average value of black hole mass uncertainties due to measurement uncertainties is 0.07 dex, while the average contribution from systematic uncertainties is 0.5 dex for broad H$\alpha$ AGN and 1.2 dex for the others. Such discrepancy comes majorly from the spread in $\alpha$.

Of the six objects previously classified as AGN, $M_{\textrm{BH}}$ has been measured for NGC 0863 and NGC 5548 using reverberation mapping. The measured values of $\log(M_{\textrm{BH}}/M_{\odot})$ are $7.57^{+0.06}_{-0.07}$ for NGC 0863 \citep{bentz18} and $7.51^{+0.23}_{-0.14}$ for NGC 5548 \citep{pancoast14}, which are in agreement with our estimates of $8.0\pm0.5$ and $7.8\pm0.5$.

\begin{figure}
\centering
\includegraphics[width=4.5in]{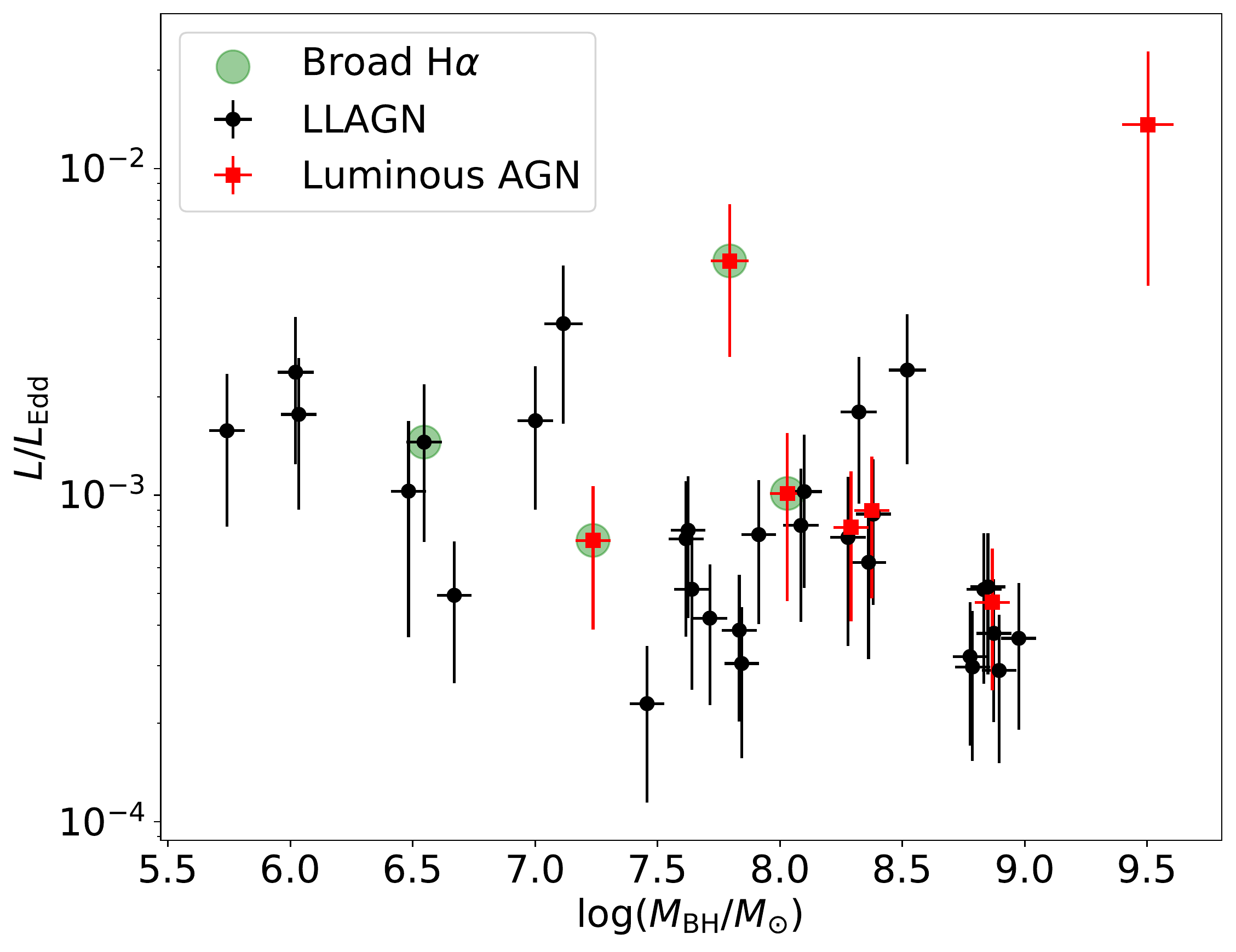}
\caption{Eddington ratio versus black hole mass for the AGN candidates, with the LLAGN/luminous AGN classification based on H$\alpha$ luminosity. The errors are purely statistical. The systematic uncertainty from converting the ASAS-SN photometry to a black hole mass is about 1.1 dex.}
\label{eddratio_plot}
\end{figure}

Given the estimated black hole mass, we can calculate the expected Eddington Luminosity, $L_{\textrm{Edd}}$. We also estimate the Eddington ratio, $L/L_{\textrm{Edd}}$, from variability. Because of its low resolution, we can only measure the variability of the AGN with ASAS-SN and not its mean flux. \citet{macleod10} find that the variability and Eddington ratio have power-law relation, $\sigma_{\textrm{XS}}\propto(L/L_{\textrm{Edd}})^{-0.23}$. This can be scaled to $L/L_{\textrm{Edd}}\sim38.0(\Delta F/F_{\textrm{Edd}})^{1.3}$, assuming the fractional variability is small. Most of our AGN candidates have $\Delta F/F_{\textrm{Edd}}$ in order of $10^{-4}\sim10^{-3}$. In this regime, this power-law relation can be approximated as $L/L_{\textrm{Edd}}\sim3.0\times\Delta F/F_{\textrm{Edd}}$. From the measured luminosity and $M_{\textrm{BH}}$ calculated from reverberation mapping, \citet{pancoast14} report that the Eddington ratio of NGC 5548 is 0.017, while for the ASAS-SN light curve, the ratio of the flux variability to Eddington is $\Delta F/F_{\textrm{Edd}}=5.2\times10^{-4}$. The ratio between these two values is in order of $10^1$, so instead of scaling by $\sim$3.0, we estimate the Eddington ratio by

\begin{equation}
    \textrm{Eddington ratio} = 10 \times \frac{\Delta F}{F_{\textrm{Edd}}}.
    \label{eddratioeq}
\end{equation}
The dependence of this Eddington ratio estimate on the black hole mass is shown in Figure \ref{eddratio_plot}. The Eddington ratios of the AGN candidates are of order $10^{-4}$--$10^{-2}$, and all the LLAGN identified based on the H$\alpha$ luminosity have Eddington ratios in order of $10^{-3}$ or lower. 

\section{Discussion and Conclusions}
\label{secconclusion}


We selected 37 AGN using the ASAS-SN light curves from a sample of 1218 bright ($g<14$ mag) SDSS galaxies with spectra, although the exact number of candidates depends on analysis choices.
To determine whether an object is variable, we calculated its excess variance in magnitudes. We also tried using the normalized excess variance, $\sigma_{\mathrm{NXS}}=\sigma_{\mathrm{XS}} / \overline{x}^2$, of the flux. This resulted in more objects being considered variable (203 in the $V$ band and 322 in the $g$ band with 103 in both), but none of the additional variable sources had AGN-like structure functions.

We also analyzed the $V$- and $g$-band light curves separately and found more AGN candidates in the $V$-band data, probably because of its longer time span (2000 vs. 600 days). 
\citet{baldassare20} also found that the number of variability selected AGN increases with the length of the light curve. 
Combining the $V$ and $g$ light curves is difficult because of the need to cross-calibrate them to 0.01 mag or better.


Excluding the seven luminous AGN in our sample, we found 30 LLAGN and constrain the LLAGN fraction to be 2.5\% down to a median Eddington ratio of $2\times10^{-3}$. However, our AGN fraction estimate for the luminous AGN (1\%) can be incomplete because known nearby luminous AGN can be excluded as SDSS spectroscopic targets \citep{strauss02}. In addition, variability selection can be insensitive to Type II AGN where there is no obvious emission from the accretion disk.

Compared to recent variability selected AGN studies, \citet{baldassare20} have the most consistent results with our study.
\citet{baldassare20} measured the light curves of galaxies using Palomar Transient Factory $R$-band data and compared them to the DRW model to identify AGN. From the sample of $\sim50,000$ galaxies, they identified $\sim1\%$ as variability selected AGN. When they analyzed the available spectra, they found that approximately 60\% of their candidates were in the star-forming region on the BPT diagram. They also found that those in the star-forming region tended to have lower stellar mass. 
They also used the DRW parameters to estimate the black hole mass and found that most of their AGN candidates have $M_{\textrm{BH}}\sim10^6 M_{\odot}$--$10^8 M_{\odot}$. 

\citet{elmer20} selected AGN based on their long-term near-infrared variability in the UKIRT Infrared Deep Sky Survey Ultra Deep Survey. They selected objects with $\chi^2>30$ for being variable and with AGN-like colors and morphologies. They found 393 variability selected AGN from 152,682 sources or 0.3\%, an order of magnitude lower than our fraction, but the typical UKIRT light curves have much lower cadence, at most seven epochs spanning over 7 yr.
Many variability selection studies do not use galaxies as the parent sample and so cannot measure the AGN fraction \citep[e.g.,][]{butler11,macleod11,kumar15,sanchezsaez19}. 

Other studies have found that the LLAGN fraction of still lower Eddington ratios is significantly higher. \citet{ho97apj} report that approximately 43\% of nearby galaxies from the Palomar Survey are active, down to Eddington ratios of $10^{-5}$--$10^{-6}$, with over 85\% of them having H$\alpha$ luminosities less than $10^{40}$ erg s$^{-1}$. 
\citet{Kewley_06} analyzed SDSS spectra of over 80,000 emission-line galaxies and identified 17\% of them as AGN. From the same AGN sample, \citet{kauffmann03} use the \OIII\hspace{1pt} luminosity to identify LLAGN, down to $\log (L_{\textrm{\OIII}}/L_{\odot})=7$. With this limit, over $80\%$ of the sample was classified as LLAGN. 

To summarize, we used variability to search 1218 bright ($g<14$ mag) SDSS galaxies with spectra for AGN activity and found that

\begin{enumerate}
    \item Approximately 3\% of the galaxies are AGN candidates showing variability and a red noise structure function. 
    \item Of the AGN candidates, 60--80\% are confirmed as AGN using SDSS spectra. About 10\% are broad-line AGN and 50--70\% lie in the AGN region of an emission line diagnostic plane.
    \item The LLAGN fraction is 2\% down to an Eddington ratio of $\sim 10^{-3}$. The luminous AGN fraction may be incomplete because it can be excluded from the SDSS spectroscopic sample.
\end{enumerate}

The strength of this method is that it requires only photometric data, and large nonspectroscopic surveys can be used to construct the parent galaxy samples, such as SDSS, Pan-STARRS, DES, 2MASS, and WISE, and LSST/Rubin in the near future for fainter and more distant galaxies. With more and deeper data, we can constrain the AGN fraction among galaxies more accurately in the local universe than using the ASAS-SN survey.  Furthermore, since the photometric galaxy sample is relatively unbiased compared to the spectroscopic sample, the variability selection can constrain the AGN fraction to certain Eddington limits including the luminous AGN and LLAGN.

\begin{acknowledgments}
We thank the anonymous referee for helpful comments and suggestions.  We acknowledge the financial support from the NASA ADAP program NNX17AF26G. H.F. acknowledges support from NSF grants AST-1614326 and AST-2103251. C.S.K. and K.Z.S. are supported by NSF grants AST-1908570 and AST-1814440. We thank Las Cumbres Observatory and its staff for their continued support of ASAS-SN. ASAS-SN is funded in part by the Gordon and Betty Moore Foundation through grants GBMF5490 and GBMF10501 to the Ohio State University and also funded in part by the Alfred P. Sloan Foundation grant G-2021-14192.
\end{acknowledgments}

\bibliographystyle{apj}
\bibliography{reference}

\end{document}